\documentclass[useAMS,usenatbib]{mn2e}
\usepackage{graphicx}

\title[CCD Photometry of M2.RR Lyrae physical parameters]{CCD Photometry of
 the globular cluster M2. RR Lyrae physical parameters and new variables\thanks{Based on
  observations collected at the Observatorio del Teide, Tenerife, Spain, and the Indian
   Astrophysical Observatory, Hanle, India.}}
\author[C. L\'azaro et al.]{C. L\'azaro$^{1,2}$\thanks{E-mail:
clh@ll.iac.es}, A. Arellano Ferro$^{3}$\thanks{E-mail:
armando@astroscu.unam.mx}, M.J. Ar\'evalo$^{1,2}$, D.M. Bramich$^{4}$,
\newauthor{S. Giridhar$^{5}$, E. Poretti$^{6}$}\\
$^{1}$Departamento de Astrof\1sica, Universidad de La Laguna.\\
$^{2}$Instituto de Astrof\1sica de Canarias.\\
$^{3}$Instituto de Astronom\1a, Universidad Nacional Aut\'onoma de M\'exico.\\
$^{4}$Institute of Astronomy, University of Cambridge, Madingley Road, 
Cambridge, CB3 0HA, UK\\
$^{5}$Indian Institute of Astrophysics, Koramangala 560034, Bangalore, India\\
$^{6}$INAF--Osservatorio Astronomico di Brera, via E. Bianchi 46, I-23807 Merate, Italy}

\begin{document}

\date{Accepted . Received ; in original form }

\pagerange{\pageref{firstpage}--\pageref{lastpage}} \pubyear{2002}

\maketitle

\label{firstpage}

\begin{abstract}
We report the results of CCD $V$ and $R$ photometry of the RR Lyrae stars in 
M2. The periodicities of most variables are
revised and new ephemerides are calculated. Light curve decomposition of the RR Lyrae stars
was carried out and the corresponding mean physical parameters [Fe/H]= $-1.47$,
 $T_{\rm eff}$= 6276 K,
 log $L= 1.63 ~L_\odot$ and $M_V = 0.71$ from nine RRab and [Fe/H]= $-1.61$, $M= 0.54 ~M_{\odot}$,
  $T_{\rm eff}$ = 7215 K, log $L = 1.74 ~L_{\odot}$ and $M_V= 0.71$ from two RRc 
stars were calculated. A comparison of the radii obtained from the above luminosity and
 temperature
with predicted radii from nonlinear convective models is discussed. 
The estimated mean distance to the cluster is $10.49 \pm 0.15$ kpc. These results place
M2 correctly in the general globular cluster sequences Oosterhoff type, mass, luminosity 
and temperature, all as a function of the 
 metallicity. Mean relationships for $M$, log $L/L_{\odot}$, $T_{\rm eff}$ and $M_V$ as a 
function of [Fe/H] for a family of globular clusters are offered. These trends are consistent 
with evolutionary and structural notions on the horizontal branch. 
Eight new variables are reported.
\end{abstract}
      

\section{Introduction}

RR Lyrae stars are of particular importance in the age and distance determination to ancient
 stellar systems, such as globular clusters. Their usefulness is based 
in the fact that their light curves are easily distinguishable and  their high intrinsic
 brightness allows their detection in systems of the Local Group. Their
well defined absolute magnitude allows using them as standard candles in the cosmic distance
 scale (e.g. Alcock et al. 2004 for the LMC).

The RR Lyrae stars have been used in the determination of absolute ages of globular clusters,
 by measuring the magnitude difference between the main sequence turn
 off point and the horizontal branch (HB), on which the RR Lyrae stars reside. This is a useful
  reddening free parameter that helps in tracing the early stages of the formation of our Galaxy.
   This, however, requires a proper calibration of the absolute magnitude $M_V$ for the RR Lyraes.

The pulsational behaviour of RR Lyrae stars in globular clusters has been a subject of study for 
decades since it offers insight on stellar evolution in the HB stages. 
The mean periods of fundamental 
pulsators (type RRab) divide globular clusters into two groups,
the Oosterhoff type I (OoI) with $<P_{ab}>$ smaller than 0.6 days and the Oosterhoff type II (OoII) with
$<P_{ab}>$ larger than 0.6 days 
(Oosterhoff 1939).  Recent light curve Fourier 
decomposition calculations have clearly shown that 
RR Lyrae stars in OoII clusters are, on average, more luminous, more
massive and cooler than in OoI clusters, and that these quantities are closely correlated 
with the mean cluster metallicity. For example these stars have longer periods in metal poor clusters
 than in metal rich clusters
 (e.g. Simon \& Clement 1993; Clement \& Shelton 1997; Kaluzny et al. 2000; Arellano Ferro et al. 2004;
Arellano Ferro et al. 2006 (see their Tables 8 and 9)).

On the other hand the period distribution of fundamental periods ($P_{ab}$ for RRab stars), and the
 fundamental periods calculated from the first overtone period and the period ratio
($P_f$ for RRc stars), as well as the relative number of RRc and RRab stars, provide important
insights relevant to the instability strip structure and HB structure and evolution (Castellani et
al. 2003). Nevertheless, such studies are strongly limited by bona fide completeness of the
sample of RR Lyrae pulsators in a cluster. It is evident from recent works that 
a substantial number of RR Lyrae stars can be discovered when new image-subtraction techniques
(Alard 2000; Alard \& Lupton 1998; Bond et al. 2001; Bramich et al. 2005)
are applied to CCD images of globular clusters (e.g. Lee \& Carney 1999a; Kaluzny et al. 2001;
Clementini et al. 2004). 

The latest CCD study of M2 (NGC 7089) was published by Lee \& Carney (1999a) (hereinafter LC99).
 These authors reported 13
new RR Lyraes, for a total of 34 variables in the cluster, 30 of which are RR Lyrae stars.
 They calculated the ratio 
n(c)/n(ab+c) = 0.40, which
approaches the mean typical value of 0.44 in OoII clusters. Hence, not many undiscovered RR Lyraes are
expected. However, since these authors did not perform a light curve decomposition analysis, 
we decided to
supplement their $BV$ data with new $VR$ observations and use the two data sets to refine the ephemerides,
 determine the mean physical parameters of the RR Lyrae stars from the Fourier decomposition
technique, compare the results with those of other similarly studied OoI and OoII clusters and,
in passing, search for new variables. The effort was not fruitless.

In Sect. 2 we describe the observations and data reductions. In Sect. 3 we calculate new
ephemerides, new times of maximum, discuss individual objects and report new variables. In Sect. 4
we calculate the physical parameters using the light curve decomposition method. In Sect. 5 we
discuss the results in the wider context of other globular clusters and in Sect. 6 we state our conclusions.

\section{Observations and Reductions}
\label{sec:Observations}

The observations used in the present work, performed using the Johnson $V$ and $R$ filters,
were obtained from June 6 to 19, 2002
with the 0.82~m telescope of the 
Observatorio del Teide, Tenerife, Spain. The seeing conditions were good, with  an estimated
$FWHM \simeq 1 ~$arcsec.
The detector was a 
Thompson CCD of 1024 $\times$ 1024 pixels 
of 19 square microns, these images are of approximately $7 \times 7$ arcmin$^2$.
 Exposure times were 900 sec and 500 sec in the $V$ and $R$ filters,
respectively.
 A total of 60 images in $V$ and 60 in $R$ were obtained.
  Higher resolution
images for proper star identification were acquired with the 2.0 m Himalayan Chandra Telescope (HCT)
 of the Indian Astrophysical Observatory (IAO).

Differential photometry by the image subtraction method, described
in detail by Bond et al. (2001) and Bramich et al. (2005), was performed. 
This procedure involves
the matching of a high quality reference image to each image in the time series,
 by solving for a spatially-varying convolution kernel and
differential sky background function. Difference images are constructed
via the subtraction of the convolved reference image from the time series
images. Photometry on the difference images yields differential fluxes for
each star relative to the flux from the reference image. Conversion of the
light curves to magnitudes requires an accurate measurement of the
reference flux.

We measured the stellar fluxes on the reference frame using DAOPhot
(Stetson 1987). However, the point-spread function (PSF) on the images was
highly non-gaussian and not fully modelled by the DAOPhot routine.
The reason for the poor PSF seems to be related with some optical
misaligment of the telescope at the time of our observations.
 Hence
our reference fluxes contain systematic errors that may 
affect the amplitude
of the variable star lightcurves.
Fourier decomposition and the subsequent calculation of physical parameters
 is mainly light curve shape dependent, and
we can use our light curves for that purpose. 

The instrumental magnitudes were converted to the $V$ standard system by 
using 168  standard stars in the interval $19.5 > V > 14.0$ listed by Stetson (2005)
in the M2 field.
The transformation equation had a linear form
 $M_{std} = 0.985(\pm 0.005)~M_{ins} - 3.990(\pm 0.01)$, without the need of a quadratic 
term; moreover, the standard stars span a $B-V$ range from $-3.5$ to $+5.0$
and no significant colour term was found.

We noticed that the difference in magnitudes between our data and LC99 light curves
changes from star to star, which is likely due
 to poor PSF quality. Then, we have shifted in magnitude each of our V light
  curves in order to match
 the LC99 photometry. Once the correct shift is applied, both data sets agree
very well, as can be seen in Fig. 1. Once the vertical matching was
accomplished we noticed a small dispersion along the time axis. Since our
 observations were gathered between 4 and 6 years after those of LC99,
   we decided to attempt a refinement of the periods, using both
data sets. The string-length method, frequency analysis and
   Fourier fits were used to detect and refine period values.
 The complete set of periods used below for the Fourier
decomposition are listed in Table 1, where the newly determined periods are
 presented alongside the periods from LC99. The differences in the periods are small
 but in several cases significant when phasing the light curves. The error on the
refined values of the periods is $2 \times 10^{-6}$ d.

In Figs. 1 and 2 the light curves in $V$ and $R$ for the previously known RR Lyraes
 are displayed. 
Different symbols are used for the observations of LC99 and our own as described
in the caption, which illustrates the general good agreement between the two data sets.

\begin{figure*} 
\includegraphics[width=16.cm,height=21.cm]{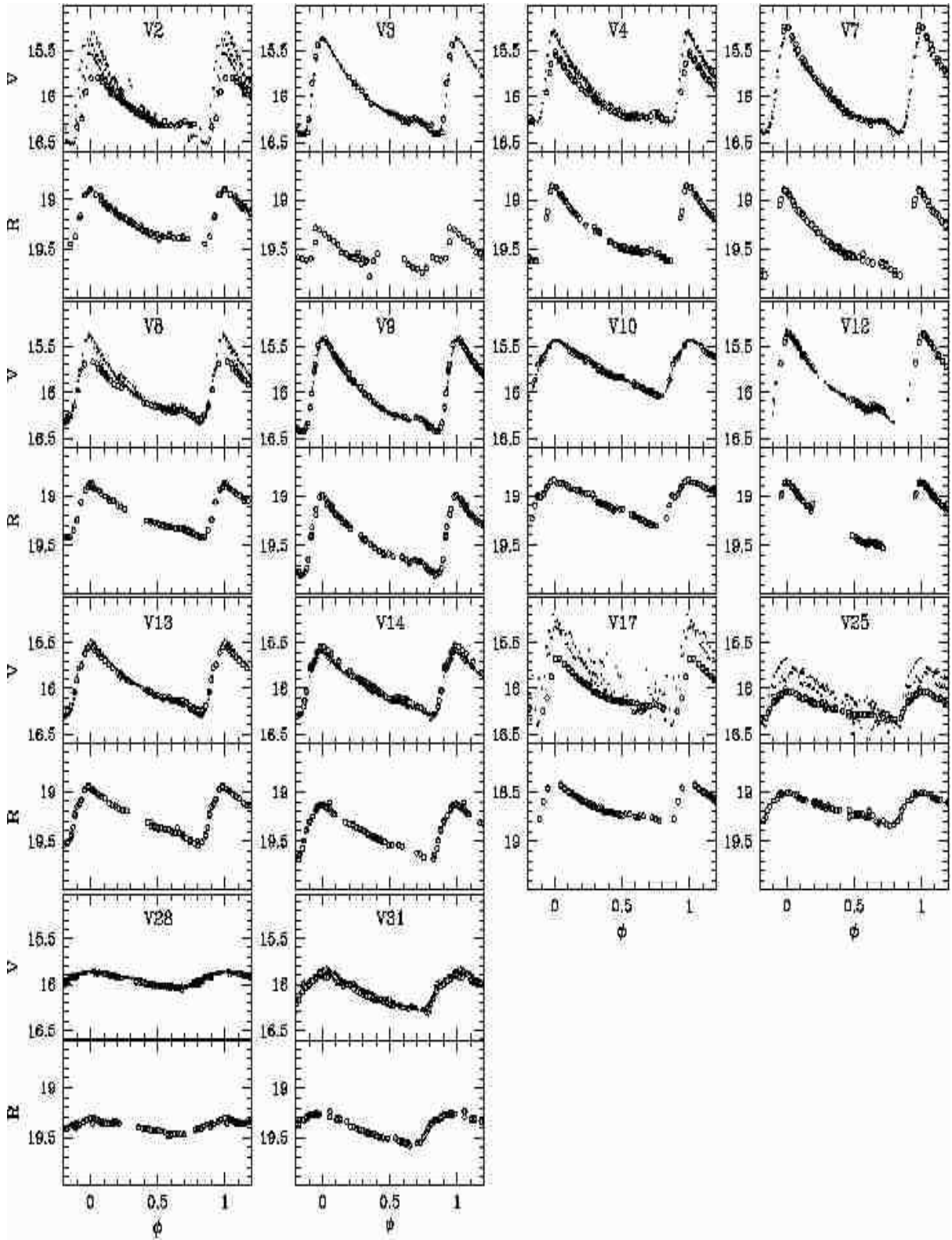}
%\vspace{4cm}
\caption{Light curves of known RRab stars in M2. They have been phased 
with the new ephemerides in Table 1. The vertical scale is the same for all the stars.
Small dots represent observations from LC99. Open circles
represent observations from the present work. }
\end{figure*}

\begin{figure*} 
\includegraphics[width=16.cm,height=11.cm]{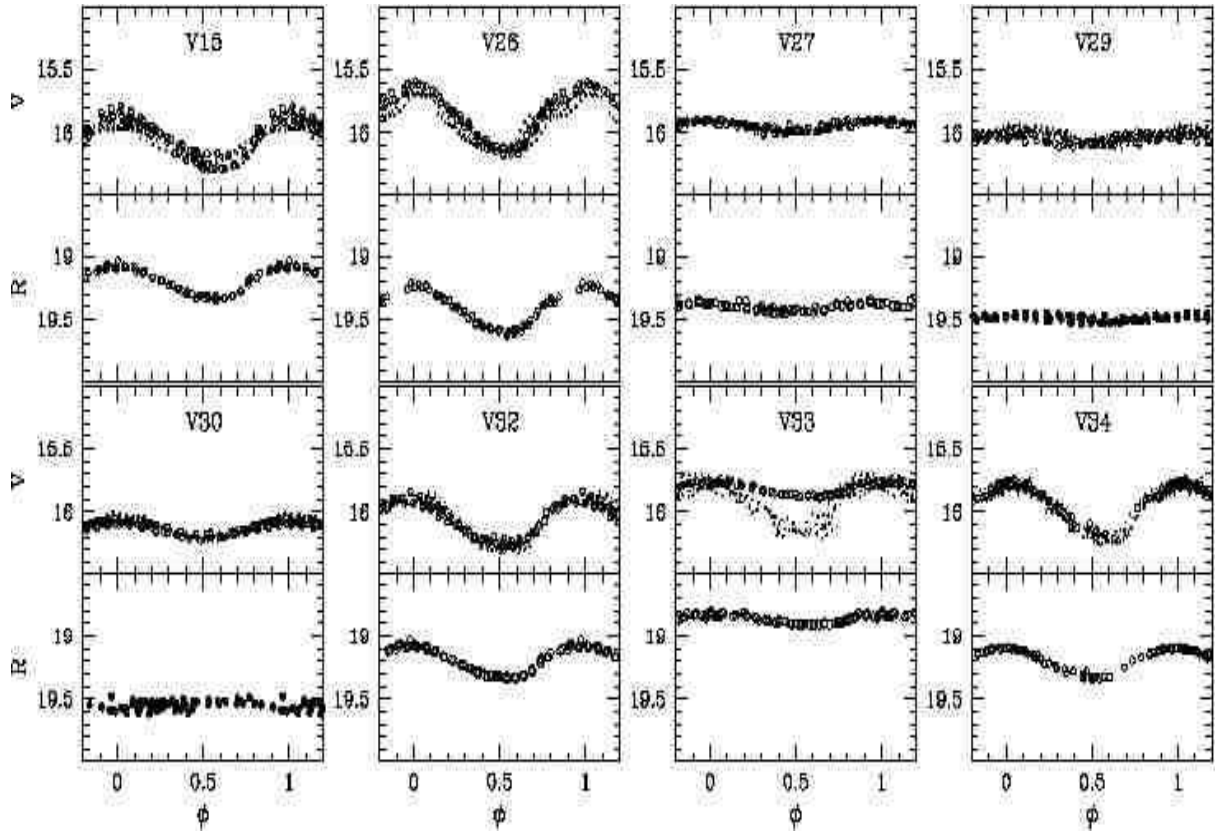}
\caption{Light curves of previously known RRc stars in M2. 
Simbols are as in Fig. 1}
\end{figure*}

\begin{table}[t!]
\footnotesize{
\begin{center}
\caption[Periodos] {\small Update of periods and epochs for the RR Lyrae stars in 
M2. \label{periodos} }
\hspace{0.01cm}
\begin{tabular}{lclll}
\\
\hline
\hline
Variable & Type & Old Period &  New Period   & New Epoch       \\
         &      &(days)      & (days)        & (+ 240 0000)\\
\hline
\hline
V2  & ab & 0.5278619 & 0.527840 & 52445.561 \\
V3  & ab & 0.6197084 & 0.619713 & 52445.165 \\
V4  & ab & 0.5642512 & 0.564243 & 52445.629  \\
V7  & ab & 0.5948665 & 0.594868 & 52445.347  \\
V8  & ab & 0.6437059 & 0.643690 & 52445.287 \\
V9  & ab & 0.6092938 & 0.609295 & 52445.582 \\
V10 & ab & 0.8757413 & 0.875744 & 52445.090  \\
V12 & ab & 0.6656063 & 0.665607 & 52445.248   \\
V13 & ab & 0.7066260 & 0.706619 & 52445.074   \\
V14 & ab & 0.6937767 & 0.693788 & 52445.442   \\
V15 & c  & 0.3007852 & 0.300785 & 52445.586 \\
V17 & ab & 0.6364715 & 0.636444 & 52445.474  \\
V25 & ab & 0.7287186 & 0.728720 & 52445.466  \\
V26 & c  & 0.4195213 & 0.412376 & 52445.461  \\
V27 & c  & 0.3141578 & 0.314158 & 52445.473  \\
V28 & ab & 0.8237775 & 0.823797 & 52445.210  \\
V30 & c  & 0.2728723 & 0.272871 & 52445.582  \\
V31 & ab & 0.7887144 & 0.788715 & 52445.005   \\
V32 & c  & 0.3619382 & 0.367021 & 52445.655 \\
V34 & c  & 0.3914157 & 0.391414 & 52445.639  \\
\hline
\hline
\end{tabular}
\end{center}
}
\end{table}

\section{Discussion of individual objects and new variables}

\subsection{Comments on individual variables}

For stars of steady amplitude (V3, V7, V9, V10, V12, V13, V27, V31 and V34), the agreement between the light curves of 
LC99 and those in the present work is excellent. No further 
discussion is needed in these cases. Other stars deserve some comments.

V2. This star is labelled as a Blazhko variable in the catalogue of Clement (2002).
 The appearance of the light curve in Fig. 1 indeed suggests that
the effect is present. The detailed plot of the light 
curve displayed in Fig. 3 
confirms the Blazhko nature of the variable and allows an
estimate of the magnitude variation of the maximum at different epochs (Table 2).
With only five measurements of the maximum light momenta it is difficult to characterize the
Blazhko periodicity. Nevertheless, the frequency analysis
     detected a secondary peak at 1.8870 c/d, very close to the main
     peak at 1.8945 c/d, which suggest a Blazhko periodicity of $\sim$133 d, however, the value of the secondary peak is
     not fully reliable owing to the poor phase coverage of the Blazhko cycle.

\begin{figure} 
\includegraphics[width=84mm]{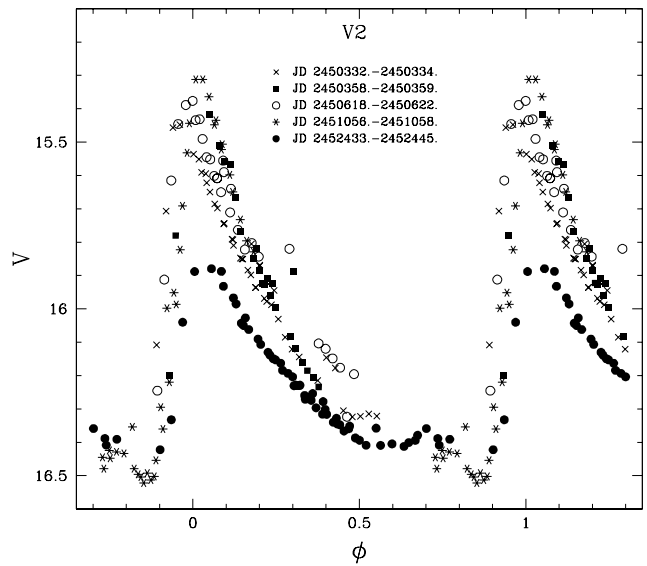}
\caption{Blazhko effect in V2.}
\end{figure}

\begin{table}
\footnotesize{
\begin{center}
\caption[Periodos y epocas] {V magnitude of maximum light variations in V2.}
\hspace{0.01cm}
\begin{tabular}{lc}
\\
\hline
\hline
$V_{max}$&HJD\\
\hline
\hline
15.532 &245 0333.667\\
15.309 &245 0358.476\\
15.374 &245 0622.921\\
15.309 &245 1057.877\\
15.878 &245 2436.609\\

\hline
\hline
\end{tabular}
\end{center}
}
\end{table}

V4, V8, V14. These stars have not been noticed before as having the Blazhko effect.
 However, the LC99 light curves show
       some cycle--to--cycle variations. Moreover, our data have
       a different amplitude in all three cases, thus, the possibility of
       a long--term Blazhko effect is just plausible, but not
       definitely proven. 

V15. The light curve in our Fig. 2 and in Fig.5 of LC99, make 
     the star suspected of having cycle--to--cycle variations. Since 
     our analysis did not detect a radial double--mode pulsation, the
     only possibility is a Blazhko effect. Even though the time series 
     has large gaps, we can tentatively explain the light curve with a
     close doublet of frequencies ($f_1$=3.3246 and $f_2$=3.2101 c/d or a
      Blazhko period of $\sim$ 8.7d).
     More observations are required to define better this doublet, since
     it seems too much separated (Moskalik \& Poretti 2003).

V17 and V25. Our light curves display a lower amplitude than in LC99, but they
    are much more accurate and rule out the possibility of a double--mode
    pulsation or a Blazhko effect. Indeed, the inspection of these stars
    on the reference image reveal that they are blends, and hence our
    amplitudes are underestimated.

V26. This star is suspected of being a double mode star in the list of Clement (2002). However
 when subdividing the data sets of LC99 and our own
into subsets, the resulting light curves look very stable although they show considerable
 time shifts ranging between 0.004~d and 0.062~d. Once the subsets are corrected by these
  shifts, a clean sinusoidal lightcurve is obtained with a period of 0.412376 days. Thus the
   star is not a double mode pulsator and the shifts can possibly be explained as light time
    effects if this RRc is a member of a binary system. More observations would be needed
to confirm this possibility.

V28. This variable seems to be rather peculiar,
 displaying a lower amplitude light curve than the other RRab stars. It is the longest
  period RR Lyrae variable in M2 (see Table 1), and its physical parameters are clearly
  discordant with those of the other RRab stars, with a particularly low 
$T_{\rm eff}$  and large luminosity and radius. The possibility of an unseen blend
  affecting its light curve can not be discarded. The star could be 
     an anomalous Cepheid, but its mean $V$ magnitude ($V$=15.945) 
     is quite normal for an RR Lyrae variable. Owing to this peculiar behaviour,
 we didn't include its physical parameters in Tab.~6.

V30. This star is also suspected of being a double mode star in the list of Clement (2002).
 However we find it to be monoperiodic with a period of 0.272871 days. The scatter in the
  observations of both LC99 and our own is a bit large.

V32. The LC99 data show some scatter but no traces of a second periodicity have been
 found performing a frequency 
      analysis. However, when splitting the time series into different
      subsets, we clearly evidenced the same light curve displaced in 
      time. Therefore, a light--time effect is plausible for this star. The
       light curve displayed in Fig. 2 has been corrected of this effect.

\subsection{Negative detections}

V16. This star is reported as variable in the catalogue of 
Clement (2002) and it is identified in the map of LC99. 
However we find no star in the position marked by these authors nor 
 variablility in nearby stars.  Therefore we have not included this star in
  Fig. 1 or in the subsequent analysis. 

V18-21. These variables were not in the field of our CCD images.

V22-24. These stars, convincingly variables as reported by LC99, 
were identified in our images, but no significant light variations were detected in either
the $V$ or $R$ filters despite not being in particularly crowded regions.

V29, V33. These stars are a blend in our images and the resulting
    amplitudes are very small. Our data are not useful to improve LC99
    results.

\subsection{Newly discovered variables}

We have detected clear light variations in 8 stars in the field of M2, not
previously reported as variables. The corresponding identifications are
 shown in Figs. 4 and 5.
 In Table 3, we report the period and epoch
of maximum light, and the Bailey's type for each new variable.  The error on the
 periods of the new variables is $4 \times 10^{-4}$ d.

The light curves of the new variables are shown in Fig. 6. Due to the problem with the
PSF of our images described in Sect. 2, and the fact that 
these new variables are all blends to some degree (see Fig. 5), we cannot assign a
 standard magnitude for the new variables. The light curves are therefore displayed
  as differential fluxes in units of ADU/sec.
Fourier decomposition and the subsequent calculation of physical parameters
are mainly light curve shape dependent, and we can use the light curves for some of the 
new variables for that purpose. The new variables are:

\begin{table}
\footnotesize{
\begin{center}
\caption[Nuevas variables] {\small Bailey's types, $(\alpha, \delta) (2000.)$
coordinates, periods and epochs for the newly detected variables. \label{nuevasvar} }
\hspace{0.01cm}
\begin{tabular}{lcllllll}
\\
\hline
\hline
Variable & Type & ~~~~$\alpha$ & ~~~$\delta$ & Period & Epoch  \\
         &      & (h ~m ~s) & ($^o$ ~$^{'}$ ~$^{"}$) & (days) & (+ 240 0000)  \\
\hline
\hline
V35  & c  & 21 33 28.0  & $-0 ~47 ~31$ &  0.32557 &  52445.517  \\
V36  & c  & 21 33 30.7  & $-0 ~49 ~12$ &  0.27078 &  52445.562  \\
V37  & ab & 21 33 26.0  & $-0 ~49 ~18$ &  0.56668 &  52445.709  \\
V38  & ab & 21 33 31.1  & $-0 ~49 ~23$ &  0.80735 &  52445.519  \\
V39  & ab & 21 33 27.3  & $-0 ~50 ~06$ &  0.60781 &  52445.467   \\
V40  & ab & 21 33 25.6  & $-0 ~49 ~15$ &  0.75173 &  52445.447  \\
V41  & ab & 21 33 28.0  & $-0 ~49 ~24$ &  0.60532 &  52445.663  \\
V42  & c  & 21 33 28.3  & $-0 ~49 ~51$ &  0.32801 &  52445.497  \\
\hline
\hline
\end{tabular}
\end{center}
}
\end{table}

\begin{figure*}
\begin{center}
\includegraphics[width=13.cm,height=10.cm]{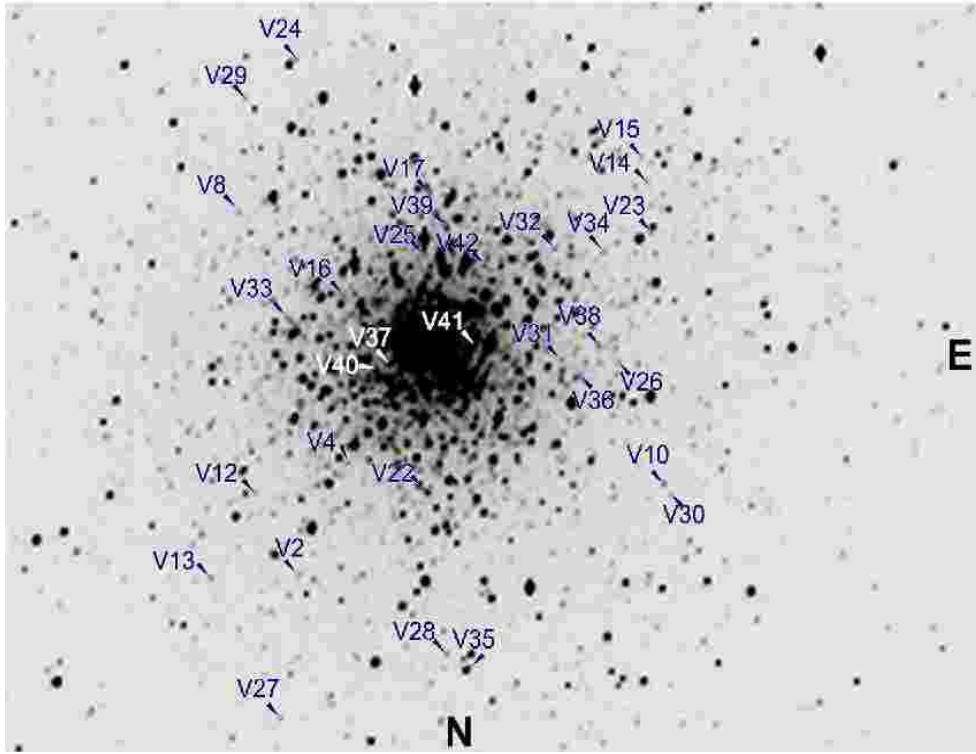}
\caption{Variables in M2. Individual image stamps for the new variables
 can be found in Fig. 5. The image was obtained at the 2.0m HCT of the IAO.
  The size of the image is approximately $6 \times 6$ arcmin$^2$.}
\end{center}
\end{figure*}

\begin{figure*}
\begin{center}
\includegraphics[width=11.cm,height=11.cm]{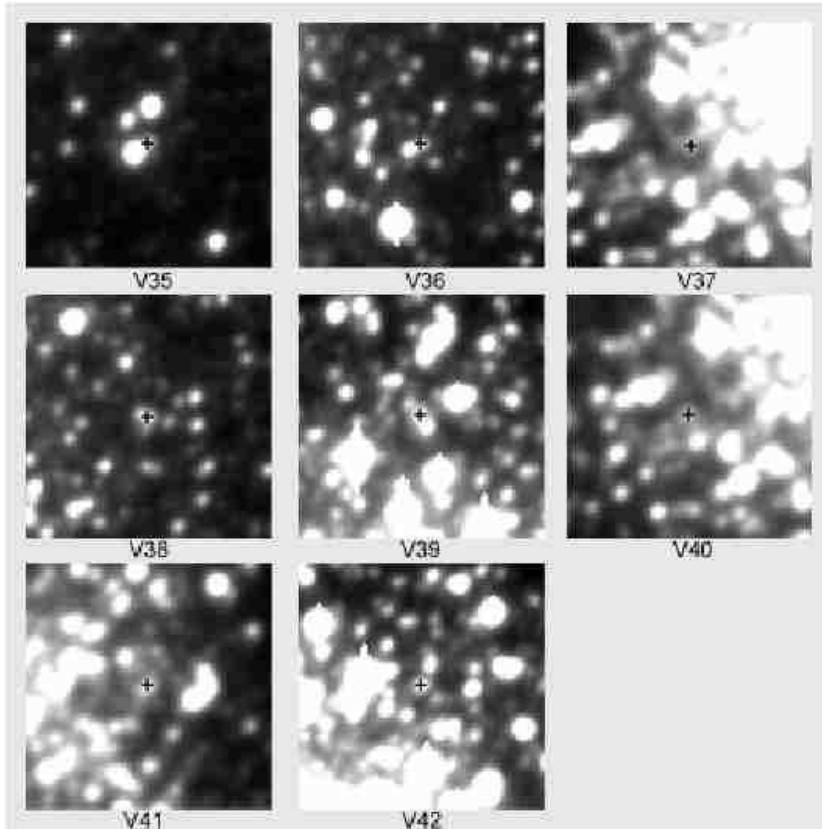}
\caption{Detailed identifications of new variables in M2. The size of the
 stamps is 17 $\times$ 17 arc seconds. As in Fig. 4 north is down and east is to the right.}
\end{center}
\end{figure*}

V35. This RRc variable appears in the V images as a blend with another 
brighter 
star by about two magnitudes, and it is not detected in the R images.
 This star will not be considered for physical parameter calculations. 

V36 and V42. These are new RRc variables. 

V37, V38, V39, and V40. These stars show clear light curves of the RRab type. 
V37 is not detected in the R images.

V41. This is a strong blend not resolved in our images. The variability in 
the difference images has a small offset relative to the centre of the 
contaminating brighter star. The variable is of type RRab.

\begin{figure*} 
\includegraphics[width=16.cm,height=11.cm]{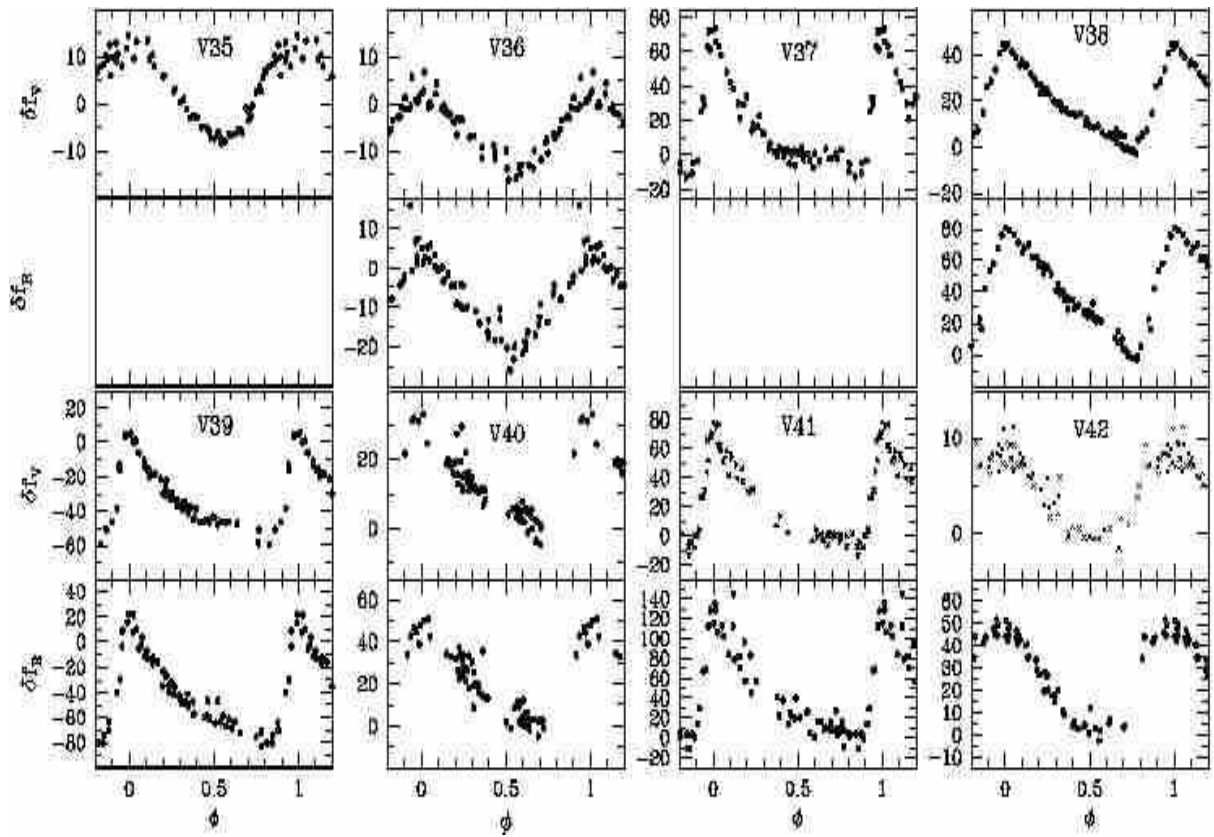}
\caption{Light curves of the new variables. For identifications refer 
to Figs. 5 and 6. The vertical scales are  differential fluxes in ADU/sec. Thus the amplitudes 
are artificial and not comparable between different filters and stars.}
\end{figure*}

\section{Fourier light curves decomposition and physical parameters}

 The mathematical representation of the light curves is of the  form:

\begin{equation}
m(t) = A_o ~+~ \sum_{k=1}^{N}{A_k ~cos~( {2\pi \over P}~k~(t-E) ~+~ \phi_k ) }
\end{equation}

\noindent
where $m(t)$ are magnitudes at time $t$, $P$ the period and $E$ the epoch. A linear
minimization routine is used to fit the data with the Fourier series model, deriving
the best fit values of $E$ and of the amplitudes $A_k$ and phases $\phi_k$ of
 the sinusoidal components. 
   
The fits are not shown in Figs. 1, 2 and 6 for simplicity and clarity. 
From the amplitudes and phases of the harmonics in eq. 1, the Fourier parameters, 
defined as $\phi_{ij} = j\phi_{i} - i\phi_{j}$, and $R_{ij} = A_{i}/A_{j}$, 
were calculated. 
The mean magnitudes $A_0$, and the Fourier light curve fitting parameters of the individual RRab and
 RRc type stars in
$V$ are listed in Table 4. The mean dispersion about the fits is about  $\pm 0.022$ mag.
These parameters will be used below to estimate the physical parameters of the stars.

The realization that light curve shapes and physical parameters are related, traces back to the 
paper of Walraven (1953). The relation between the pulsation of RR Lyrae variables and 
their physical parameters is obtained from hydrodynamic pulsation models of defined physical parameters,
 which are used to generate theoretical light curves. These curves are then fitted to obtain 
the corresponding Fourier parameters. Simon \& Clement (1993) used
 hydrodynamic pulsation models to calibrate equations for the effective temperature $T_{\rm eff}$, 
 a helium content parameter $Y$,
the stellar mass $M$ and the luminosity $L$, in terms of the period and Fourier parameter 
$\phi_{31}$ for RR Lyrae stars of the Bailey's type RRc (see eqs. 2, 3, 4 and 5).
Recently, Morgan {\it et al.} (2005) have found several [Fe/H] - log~P - $\phi_{31}$ relations for
RRc variables, and we have adopted their eq. 5 (see eq. 6 below), with a quoted
standard deviation of $\pm 0.21$, to
derive the [Fe/H] values given in Table 5. $M_V$ for the RRc stars can be estimated
 through the calibration given by Kov\'acs (1998) (eq. 7). For the sake of clarity, all the equations
  used are listed below:

%   Relaciones utilizadas para el tipo RRc :	
% (phi21 de serie senos, phi31 de serie cosenos )			'

%C Masa, log(M):
\begin{equation}
log~M/M_\odot = 0.52~log~P ~-~ 0.11~\phi^{(c)}_{31} ~+~ 0.39
\end{equation}

%C Luminosidad, log(L):
\begin{equation}
	log~L/L_\odot = 1.04~log~P - 0.058~\phi^{(c)}_{31} ~+~ 2.41
\end{equation}

%C Temperatura, log(Tef) ( de combinar la ec. de M con la ec. 2 de su articulo. Usada por
%C Alcock et al. (2004):
\begin{equation}
	log T_{\rm eff} = 3.775 ~-~ 0.1452~log~P ~+~ 0.0056~\phi^{(c)}_{31}
\end{equation}
$$	log~Y = -20.26 ~+~ 4.935~log~T_{\rm eff} ~-~ 0.2638~log~M/M_\odot ~+~ $$
\begin{equation}
~~~~~~~	~+~ 0.3318~log~L/L_\odot
\end{equation}
$$ [Fe/H] = 3.702~(log~P)^2 ~+~ 0.124~(\phi^{(c)}_{31})^2 ~-~ 0.845~\phi^{(c)}_{31} ~-~ $$
\begin{equation}
~~~~~~~	~-~ 1.023~\phi^{(c)}_{31}~log~P ~-~ 2.620 	
\end{equation}
\begin{equation}
M_V = 1.261 ~-~ 0.961~P ~-~ 0.044~\phi^{(s)}_{21} ~-~ 4.447~A_4  	
\end{equation}

For the RRab stars we have used the eq. 3 of Jurcsik \& Kov\'acs (1996) for [Fe/H] (our eq. 8),
eq. 2 in Kov\'acs \& Jurcsik (1996) for $M_V$ (our eq. 9), and eqs. 5 and 11 in Jurcsik (1998)
for $(V - K)_o$ and log $T_{\rm eff}$ (our eqs. 10 and 11). The explicit formulae are:

\begin{equation}
	[Fe/H] = -5.038 ~-~ 5.394~P ~+~ 1.345~\phi^{(s)}_{31}
\end{equation}
\begin{equation}
	M_V= 1.221 ~-~ 1.396~P ~-~ 0.477~A_1 ~+~ 0.103~\phi^{(s)}_{31}
\end{equation}
$$ (V - K)_o= 1.585 ~+~ 1.257~P ~-~ 0.273~A_1 ~-~ 0.234~\phi^{(s)}_{31} ~+~ $$
\begin{equation}
~~~~~~~ ~+~ 0.062~\phi^{(s)}_{41}
\end{equation}
\begin{equation}
	log~T_{\rm eff}= 3.9291 ~-~ 0.1112~(V - K)_o ~-~ 0.0032~[Fe/H] ~~~~
\end{equation}

The luminosity is derived through $M_V$ and the bolometric correction $BC$,
 adopting the relation of Sandage \& Cacciari (1990):
\begin{equation}
 BC = 0.06~[Fe/H] ~+~ 0.06 
\end{equation}
With:
\begin{equation}
M_{bol} = M_V ~+~ BC ~~~~ {\rm and} ~~~~ log~L/L_\odot= -0.4 ~( M_{bol} ~-~ 4.75 ) ~~~~
\end{equation}

In the previous equations, $\phi^{(c)}_{jk}$ and $\phi^{(s)}_{jk}$ are phase shifts with
assumed cosines and sines Fourier series respectively, they are related by:
  ~~ $\phi^{(s)}_{jk} = \phi^{(c)}_{jk} - (j - k) {\pi \over 2}$.\\

A thorough discussion of the uncertainties in the physical parameters, as obtained 
from the above mentioned calibrations,
can be found in the work of Jurcsik (1998). The estimated uncertainties
for log~$L$, log $T$, log $M$ and [Fe/H] are  $\pm 0.009, \pm 0.003, \pm0.026$ and $\pm 0.14$ dex
respectively. 

We have carried out the Fourier decomposition of RRab stars using the combined data of 
LC99 and our own, except in those stars where the Blazhko effect is confirmed 
or suspected, i.e. in V2, V4, V8, V14 and V15. 
 Moreover, V31, V40 and V41 data are less reliable owing to high scatter and/or gaps in the
   phase coverage: also these
 stars have been not considered. Tables 5 and 6 list the physical parameters we
  obtained.
  
  We notice that the average value of [Fe/H] we
derived from our RRab sample ($-1.47$)
is in close agreement with  that
reported by Kovacs \& Walker (2001; [Fe/H]=$-1.43\pm0.15$). Also when
including the two RRc stars (as Kovacs \& Walker did),
the average value ([Fe/H]=$-1.50 \pm 0.17$) is within the error bars.

It must be noticed that $M_V$ derived from eqs. 7 and 9 for RRc and RRab stars respectively,
 agree very well. However the luminosities 
for RRc stars derived from eq. 3 seem to be too large when compared with those for
RRab variables (derived from eqs. 9, 12 and 13). This can be appreciated in Fig. 8
 (open circles and crosses). 
The possibility that this is the result of the temperature dependence of the bolometric correction, $BC$,
 can be ruled out as it does not explain the difference in $log~L/L_\odot$. Alternatively, if eq. 12
  is used to transform the $M_V$ values, obtained with eq. 7, into $log~L/L_\odot$, both RRc and RRab
   stars have similar luminosities (crosses in Fig. 8). Moreover, the $A_0$ values
    of the  RRc  stars reported in Table~4~(V27, V30, V32 and V34) are quite similar
     to those of the RRab stars. 
 Then, some inconsistency seems to exists between the zero points of eq. 3 and eq. 7
 for the RRc stars. This problem has been noticed and commented by Cacciari et al. (2005)
   whom opted for decreasing the magnitud scale derived from eq. 7 by 0.20 mag.
 
\begin{table}
\footnotesize{
\begin{center}
\caption[Fourier fitting parameters for V light curves] {\small Fourier 
fitting parameters for V light curves}
\hspace{0.001cm}
\begin{tabular}{lcccccc} 
\\ 
\hline 
\hline 
Star (N)&$A_0$&$A_1$&$A_4$&$\phi_{21}$&$\phi_{31}$&$\phi_{41}$\\ 
\hline 
\hline  
V3 (9)&16.007&0.362&0.084&4.03&2.05&0.17\\
V7 (8)&15.995&0.391&0.084&3.95&1.91&6.14\\
V9 (8)&16.038&0.356&0.077&3.98&1.99&0.10\\
V10 (8)&15.753&0.246&0.019&4.48&3.06&1.27\\
V12 (8)&15.957&0.349&0.075&4.17&2.29&0.64\\
V13 (6)&15.952&0.275&0.045&4.19&2.33&0.61\\
V17 (8)&16.056&0.196&0.052&3.89&1.72&5.98\\
V25 (5)&16.211&0.118&0.011&4.03&2.18&0.26\\
V27 (1)&15.954&0.047& & & &\\
V28 (2)&15.945&0.079& &4.79&   &\\
V30 (1)&16.126&0.079&   &   &   & \\
V31 (4)&16.074&0.184&0.013&4.48&3.06&1.66\\
V32 (4)&16.085&0.201&0.005&4.66&3.50&2.70\\
V34 (4)&15.983&0.213&0.013&5.38&4.50&3.23\\
V36 (2)&--&0.077& &4.36& & \\
V37 (7)&--&0.123&0.040&3.61&1.51&5.59\\
V38 (4)&--&0.107&0.008&4.39&2.63&0.89\\
V39 (6)&--&0.097&0.017&3.96&2.09&6.10\\
V41 (4)&--&0.048&0.014&4.03&2.31&0.06\\
\hline
$\langle\sigma_{RRab}\rangle$&$\pm$0.002&$\pm$0.003&$\pm$0.003&$\pm$0.07&$\pm$0.11&$\pm$0.18\\
$\langle\sigma_{RRc}\rangle$&$\pm$0.002&$\pm$0.004&$\pm$0.004&$\pm$0.24&$\pm$0.37&$\pm$0.66\\
\hline
\hline
\end{tabular}
\end{center}
N: number of harmonics used to fit the light curve.\\
$A_0$ : after shifting our data to fit the LC99 magnitude scale.\\
V17 and V25 : light curve fitting on our data only. 
}
\end{table}

\section{Discussion}

\subsection{Oosterhoff type and RR Lyrae statistics}

Including the 3 new variables the average value of the period for the RRc stars is 0.327 $\pm$ 0.044 days.
 Comparing with the mean values for seven clusters
reported by Clement \& Rowe (1999), M2 fits better among the OoI type, although considering the
 standard deviations, the difference from the value in OoII type clusters of 0.36 days, is not
  significant.
For the RRab stars, including the 5 new variables, we find an average period of 0.674
 $\pm$0.093 days which is typical of Oo II type clusters (average ~0.65 days). Therefore,
  it seems that M2 is a borderline case between OoI and OoII type clusters.

\begin{table*} 
\footnotesize{ 
\begin{center} 
\caption[Parametros estelares de las RR Lyrae del tipo  c] {\small  Physical parameters for the 
RR{\lowercase {c}} stars   \label{parametrosc} } 
\hspace{0.01cm} 
\begin{tabular}{lccccccccc} 
\\ 
\hline 
\hline 
Star & [Fe/H] & $T_{\rm eff}$ & $M_V$ & $log (L/L_\odot)$ & $Y$ & $M/M_\odot$ &
 $log (R/R_\odot)$ (LT) & $log (R/R_\odot)$ (PRZ) & D ($kpc.$) \\ 
\noalign{\smallskip} 
\hline 
\hline 
\noalign{\smallskip} 
V32 & $-1.798$ & 7202. &  0.75 & 1.754 &  0.26 &  0.60 &  0.713 &  0.710 &  10.71 ~~~~ \\
V34 & $-1.422$ & 7228. &  0.66 & 1.725 &  0.28 &  0.48 &  0.673 &  0.728 &  10.64 ~~~~ \\
\hline 
Mean  &$-1.61$ & 7215. &  0.71 &  1.74 &  0.27 &  0.54 &  0.69 &  0.68 &  10.67 \\ 
$\sigma$ &  $\pm$ 0.27 & $\pm$ 18.& $\pm$ 0.06& $\pm$ 0.02 & $\pm$ 0.01 & $\pm$ 0.08 & 
$\pm$ 0.03 & $\pm$ 0.04 & $\pm$  0.05 \\ 
\noalign{\smallskip} 
\hline 
\hline
\end{tabular} 
\end{center} 
}
\end{table*}

\begin{table*}
\footnotesize{
\begin{center} 
\caption[Parametros estelares de las RR Lyrae del tipo ab] {\small Physical parameters for the RR{\lowercase {ab}}
  stars \label{parametrosab} } 
\hspace{0.01cm} 
\begin{tabular}{lccccccc} 
\\ 
\hline 
\hline 
Star  & [Fe/H] & $T_{\rm eff}$ & $M_V$  & $log (L/L_\odot)$ &  $log (R/R_\odot)$ (LT) &
 $log (R/R_\odot)$ (PRZ) & D ($kpc.$) \\ 
\noalign{\smallskip} 
\hline 
\hline 
\noalign{\smallskip} 
V3   & $-1.398$ & 6380. &  0.72 &  1.622 &  0.729 &  0.767 &  10.49 \\
V7   & $-1.452$ & 6424. &  0.72 &  1.621 &  0.720 &  0.756 &  10.40 \\
V9   & $-1.421$ & 6384. &  0.73 &  1.618 &  0.727 &  0.762 &  10.58 \\
V10  & $-1.421$ & 6084. &  0.52 &  1.702 &  0.808 &  0.854 &  10.22 \\
V12  & $-1.323$ & 6320. &  0.68 &  1.634 &  0.744 &  0.785 &  10.40 \\
V13  & $-1.490$ & 6231. &  0.67 &  1.645 &  0.758 &  0.800 &  10.47 \\
V37  & $-1.824$ & 6287. &  0.85 &  1.579 &  0.714 &  0.744 &   ~-~ \\
V38  & $-1.630$ & 6049. &  0.64 &  1.660 &  0.790 &  0.833 &   ~-~ \\
V39  & $-1.280$ & 6331. &  0.87 &  1.561 &  0.705 &  0.762 &   ~-~ \\
\hline 
Mean & $-1.47$ &  6276. &  0.71 &  1.63 &  0.75 &  0.79 &  10.42 \\ 
$\sigma$ & $\pm$ 0.16 &  $\pm$ 124. & $\pm$ 0.10& $\pm$ 0.04& $\pm$ 0.03& $\pm$ 0.04& $\pm$0.11 \\
\noalign{\smallskip} 
\hline 
\hline 
\end{tabular} 
\end{center}  
}
\end{table*}

After LC99 reported 13 new RR Lyraes, for a total of 30 in the cluster, the ratio 
n(c)/n(ab+c) = 0.40
approached the mean typical value of 0.44 in OoII clusters and argued that not many undiscovered RR Lyraes
are expected. We have found 8 new variables and a new value n(c)/n(ab+c) = 0.41
 (negative detections excluded as variables ).
We see no reason for not finding yet more RR Lyraes in this cluster if adequate monitoring is performed.

\subsection{RR Lyrae radii}

Given $log(L/L_\odot)$ ($M_V$ for RRab) and $T_{\rm eff}$ in Tables 5 and 6 one can derive the stellar radii. 
These radii depend fully on the semi-empirical relations and the hydrodynamical models used to calculate the
 luminosity and temperature, and they are included in Tables 5 and 6 under the name $log (R/R_{\odot}) (LT)$.
  Recently 
Marconi {\it et al.} (2005) have offered Period-Radius-Metallicity (PRZ) relationships for the RR Lyrae  
based on the nonlinear convective models of Bono et al. (2003 and references therein). In these 
relations the mean radius can be obtained from the pulsation period and the metallicity parameter $Z$.
We have used these calibrations to calculate the radii of the RR Lyrae in M2
and compare them with the radii obtained independently from the light curve Fourier decomposition.
 In doing so, we have converted the [Fe/H] parameter 
into Z making use of the equation: 
$log~\rm{Z} = \rm{[Fe/H]} - 1.70 + log (0.638~f + 0.362)$ where $f$ is the
 $\alpha$-enhancement factor with respect to iron (Salaris et al. 1993) which we adopt as
  $f=1$. The radius can be estimated using the individual values of [Fe/H] in Table 6
or the average value $-1.543$. We have found that the differences in the estimated radii
are not significant. Thus we report in Tables 5 and 6 the radii
from P and Z, labelled  $log (R/R_{\odot})$ (PRZ) using the average [Fe/H]. 

\begin{figure} 
\includegraphics[width=6.cm,height=8.cm, angle=-90.]{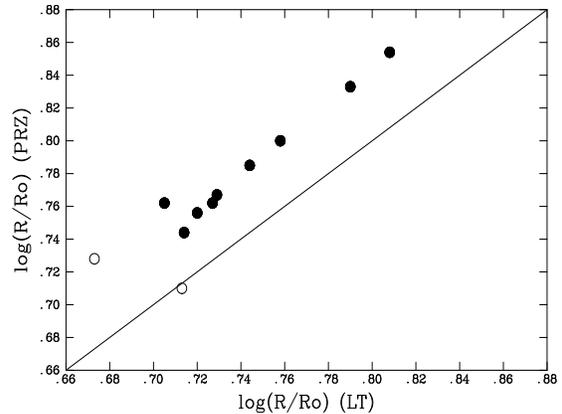}
\caption{Comparison of radii from the Fourier decomposition parameters $L$ and $T_{\rm eff}$ 
with those obtained from the Marconi {\it et al.} (2005) calibration from $P$ and $Z$.
 Solid circles represent RRab stars and open circles RRc stars}
\end{figure}

In Fig. 7 the comparison of $log (R/R_{\odot})$ (LT) and $log (R/R_{\odot})$ (PRZ)
is shown. As seen in Tables 5 and 6, and Fig. 7, the $log (R/R_{\odot})$
 (PRZ) values are systematically larger than $log (R/R_{\odot})$ (LT) by about 0.04. 
But given that both radii determinations are based on different models,
methods and calibrations, and the sensitivity of radii to the opacity and
other ingredients in the stellar models, both determinations can be considered quite
consistent.

\subsection{Distance to M2}

To calculate the distance to the cluster we have calculated the distance moduli $(A_0 - M_V)$ 
 for each star, using the absolute magnitudes
$M_V$, given in Tables 5 and 6, derived from the Fourier parameters listed in Table 4.
The total to selective extinction ratio $R= A_V/E(B - V)= 3.1$, with $E(B - V)= 0.06$ (Harris 1996),
was adopted to correct for interstellar extinction.
For the RRab stars the $M_V$ values were obtained from the calibration of
 Kov\'acs \& Jurcsik (1996),
while for the RRc stars the relation of Kovacs (1998) was used to derive $M_V$.
 
We find a mean distance of $10.42 \pm 0.12$ kpc for the RRab and
$10.67 \pm 0.05$ kpc for the RRc stars respectively. 
The uncertainty in these values is the standard deviation of the mean from individual stars.
Our mean value of the distance to M2, $10.49 \pm 0.15$ kpc,
 from eleven RRab and RRc variables, is to be compared with the  previous estimates of
 11.5 and 11.2 kpc reported by Harris (1975; 1996), respectively.

\subsection{On the evolutionary stage of the RR Lyrae stars}

From the values in Tables 5 and 6 we can place the
RRc and RRab stars in the  HR diagram in Fig. 8. 
Two versions of the instability strip are shown. 
    The vertical boundaries are the fundamental (continuous lines) and first overtone 
    (dashed lines) instability strips from Bono et al. (1995), $M= 0.65 ~M_{\odot}$,
     Y= 0.24 and Z= 0.001 .
  As shown in the figure, the RRab variables populate the fundamental mode band
  in a narrow region in $T_{\rm eff}$ near the red edge of the theoretical instability
  strip, whereas the two RRc stars are located close to the first overtone blue edge.
  The small number of RRc variables in our diagram does not allow to discuss
   the suggestion of Bono et al. (1995) that in OoII clusters the
   transition between RRc and RRab variability occurs near the first overtone
    red edge, while in the OoI clusters the transition is found closer to the
    fundamental blue edge. 
          The short continuous tilted lines are the empirical
   fundamental mode instability strip found by Jurcsik (1998) from 272 RRab stars
    of different metallicity, including variables from galactic field stars,
   many globular clusters and the Sculptor dwarf galaxy.
  It seems rather surprising that the empirical instability band is much narrower
  that the modelled one from Bono et al. (1995), even if it is defined from a
  very inhomogeneous sample of RRab variables. It is also clear that the location
   of the RRab stars in M2 follows closely the empirical strip both in slope and width. 

Two values for the luminosity of the two RRc stars are plotted in Fig. 8. The upper
 values, open circles, correspond to the values obtained from eq. 3. 
The lower values are those obtained from the Mv values from eq. 7 and transformed
 into luminosity by the eqs. 12 and 13. While the apparent and absolute magnitudes
 of the RRc stars agree with those of the RRab stars, it is clear that the luminosity 
predicted from eq. 3 is large. We shall retain these large values for further
 comparison, in section 5.5, with RRc stars
in other globular clusters studied by the Fourier decomposition technique.

The theoretical zero age horizontal branch (ZAHB) from the RRab models of Lee \& Demarque (1990)
 for two chemical mixtures,
 (Y=0.20;Z=0.0001) and (Y=0.23; Z=0.0007) are also shown in Fig. 8. These
  two ZAHB`s lie above the RRab stars. However,
  the estimated value of the relative abundance of helium for our RRc sample is Y=0.27, for
   which a model is not available. Since the luminosity of the ZAHB decreases with increasing Y,
    an extrapolation of Lee \& Demarque (1990) models to larger values of Y would make the
     ZAHB match the distribution of RRab stars.

\begin{figure} 
\includegraphics[width=84mm]{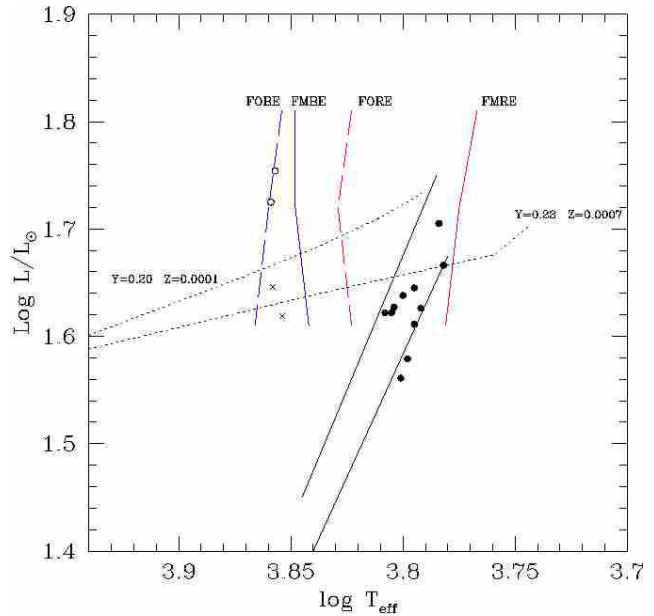}
\caption{Solid circles represent RRab stars, open circles RRc stars with $log(L/L_\odot)$
 calculated from eq. 3 and crosses RRc stars with $log(L/L_\odot)$ calculated from $M_V$.
The solid tilted lines indicate the empirical bounds found by Jurcsic (1998) 
from 272 RRab stars. The vertical boundaries are the fundamental 
(continuum lines) and first overtone (dashed lines) instability strips from 
Bono et al. (1995) for 0.65 $M/M_\odot$.
Two models of the ZAHB (Lee \& Demarque 1990) are shown and
labelled with the corresponding metallicities. An extrapolation of these models to the value 
of $Y=0.27$ (found from the RRc stars) would place the RRab stars near the ZAHB.}
\end{figure}

\subsection{Physical parameter trends in globular clusters}

The mean parameters for the RRc and RRab stars are compared for several clusters in Tables 7 and 8.
 These tables contain only those clusters for which their mean parameters have been obtained from
  the technique of RR Lyrae 
light curve Fourier decomposition and are updated versions of equivalent tables previously published
 by Arellano Ferro et al. (2004; 2006) and Kaluzny et al. (2000).

\begin{table*}
\begin{center}
\caption[Parametros RRc] {\small Mean 
physical parameters obtained from RR{\lowercase {c}} stars in globular clusters.}
\hspace{0.01cm}
\begin{tabular}{cccccccc}
\\
\hline
\hline
Cluster  & Oo type & [Fe/H] & No. of stars& $M/M_\odot$ &  $log(L/L_\odot)$ &  $T_{\rm eff}$ & $Y$ \\
\noalign{\smallskip}
\hline
\hline
\noalign{\smallskip}
NGC~6171 & I & $-0.68$& 6&0.53 &1.65 &7447&0.29 \\
NGC~4147$^1$& I &$-1.22$&9&0.55 & 1.693 & 7335& 0.28 \\
M5      &I  & $-1.25$& 7&0.58 & 1.68 &7338&0.28\\
M5$^2$      &I  &$ -1.25$&14&0.54 & 1.69 &7353&0.28\\
M3$^3$      &I  &$-1.47$& 5&0.59& 1.71 & 7315&0.27\\
NGC 6934$^4$&I  &$-1.53$& 4&0.63& 1.72 & 7300&0.27\\
{\bf M2$^5$}  & II & $-1.61$ & 2 & 0.54 & 1.739 & 7215  & 0.27\\
M9      &II&$-1.72$&1& 0.60&1.72  & 7299&0.27\\
M55$^6$ &II&$-1.90$&5&0.53& 1.75&  7193&  0.27\\
NGC~2298 &II&$-1.90$&2&0.59&1.75& 7200&0.26\\
M92$^7$ & II &$-1.87$&3&0.64&1.77&7186&0.26\\
M68    &II&$-2.03$&16&0.70&1.79&7145& 0.25\\
M15   & II&$-2.28$&6& 0.73&1.80&7136&0.25\\
M15$^8$ & II &$-2.12$&8& 0.76 & 1.81 & 7112 & 0.24\\
\noalign{\smallskip}
\hline
\hline
\end{tabular}
\end{center}
Data taken from Clement \& Shelton (1997), except:\\
1. Arellano Ferro {\it et al.} (2004), 2. Kaluzny {\it et al.} (2000),
 3. Kaluzny {\it et al.} (1998), 4. Kaluzny et al. (2001),
5. This work, 6. Olech {\it et al.} (1999),
 7. recalculated in this work from the data of Mar\1n (2002), 8. Arellano Ferro {\it et al.} (2006). 
\end{table*}

\begin{table*}
\begin{center}
\caption[Parametros RRab] {\small Mean 
physical parameters obtained from RR{\lowercase {ab}} stars in globular clusters.}\label{kaluznyab}
\hspace{0.01cm}
\begin{tabular}{cccccc}
\\
\hline
\hline
Cluster  & Oo type&No. of stars & [Fe/H]  &  $T_{\rm eff}$ & $M_V$\\
\noalign{\smallskip}
\hline
\hline
\noalign{\smallskip}
NGC~6171$^1$ & I  &3& $-0.91$ &6619&0.85   \\
NGC~4147$^2$& I  &5& $-1.22$ &6633&0.80   \\
NGC1851$^3$ & I  &7&$-1.22$  &6494&0.80    \\
M5$^4$       & I  &26&$-1.23$ & 6465&0.81  \\
M3$^5$       & I  &17&$-1.42$ & 6438&0.78  \\
NGC 6934$^6$& I& 24&$-1.53$& 6450&0.81 \\
M55$^7$  & II &5&$-1.48$ & 6352&0.71  \\
{\bf M2$^8$}& II & 9 & $-1.47$ & 6276 & 0.71  \\
M92$^9$  & II &5&$-1.87$ & 6160&0.67 \\
M15$^{10}$  & II &11&$-1.87$ & 6237&0.67 \\
\noalign{\smallskip}
\noalign{\smallskip}
\hline
\hline
\end{tabular}
\end{center}
1. Clement \& Shelton (1997), 2. Arellano Ferro {\it et al.} (2004), 3. Walker (1999),
 4. Kaluzny {\it et al.} (2000), 5. Kaluzny {\it et al.} (1998), 6. Kaluzny et al. (2001), 7. Olech {\it et al.} (1999), 8. This work, 
  9. recalculated in this work from the data of Mar\1n (2002),
 10. Arellano Ferro {\it et al.} (2006). 
\end{table*}

\begin{figure} 
\includegraphics[width=84mm,height=9.cm]{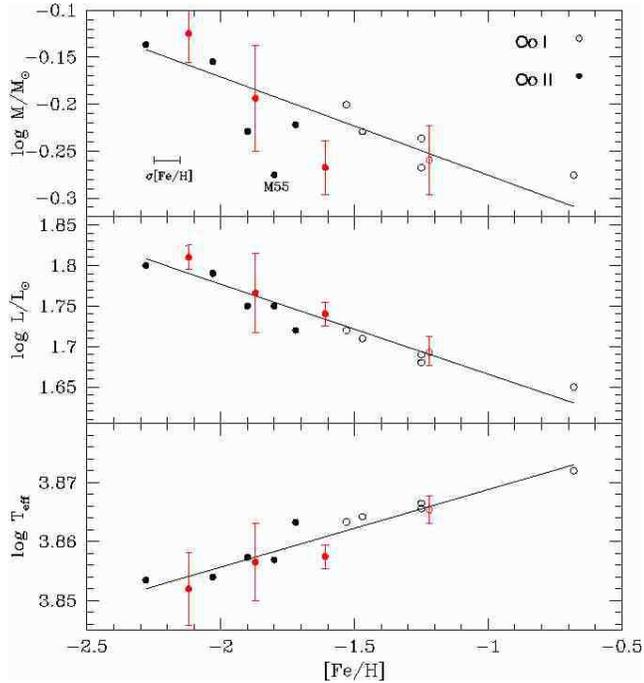}
\caption{General trends of relevant physical parameters in globular clusters
as a function of metallicity. All these parameters have been calculated by 
the RRc stars light curve Fourier decomposition technique. The error bars are the standard deviation of the
 mean divided by the square root of the number of stars included in each cluster. The horizontal error bar
  in the metallicity is a mean
uncertainty calculated using the expression from Jurcsik \& Kov\'acs (1996), their eq. 4.
 Error bars have only been calculated for those clusters studied by our team. Label dot has not been included in the fit.}
\end{figure}

\begin{figure} 
\includegraphics[width=84mm,height=6.cm]{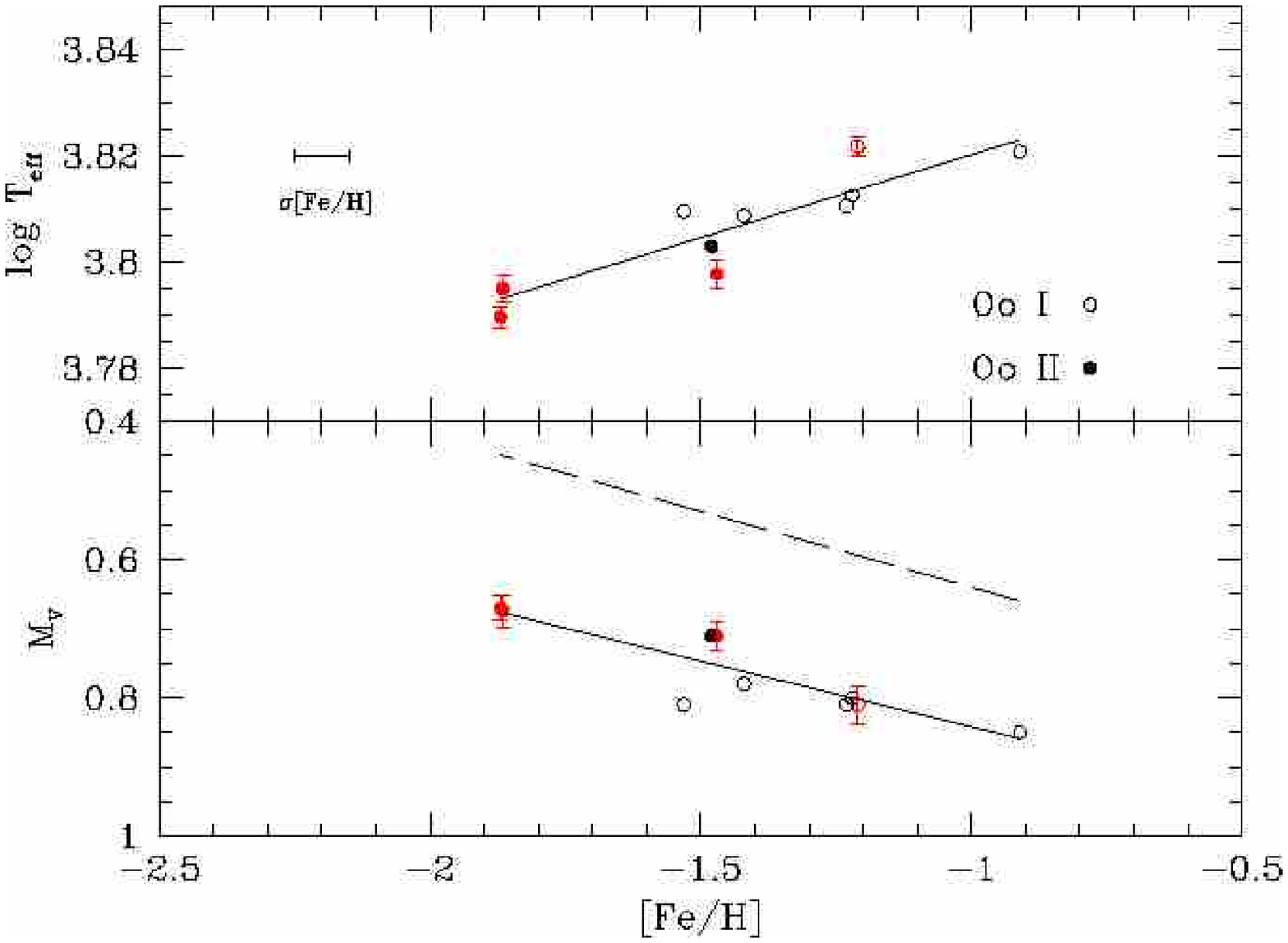}
\caption{Same as Fig. 9 but for the RRab stars. The dashed line corresponds to eq. 19.
 This indicates that eq. 3 makes the RRc stars to appear brighter than RRab stars by about 0.20 mag. See text in section 4 for discussion.}
\end{figure}

Figs. 9 and 10 show the trends of physical parameters as a function of metallicity
for RRc and RRab stars respectively. The dependences on metallicity are clear in all cases,
especially for $log L/L_{\odot}$ and $T_{\rm eff}$. The straight lines in the figures are the least square
  fits to the point distribution. The relationships representing the lines are:\\

For the RRc stars:

\begin{equation}
log~M/M_{\odot} = - ( 0.105 \pm 0.019 )~\rm{[Fe/H]} - ( 0.381 \pm 0.032 ) 
\end{equation}
\begin{equation}
log~L/L_{\odot} = - ( 0.111 \pm 0.009 )~\rm{[Fe/H]} + ( 1.554 \pm 0.016 ) 
\end{equation}
\begin{equation}
log~T_{\rm eff} = + ( 0.013 \pm 0.001 )~\rm{[Fe/H]} + ( 3.882 \pm 0.002 )
\end{equation}

The uncertainties in $log M/M_{\odot}$, $log L/L_{\odot}$, and $log T_{\rm eff}$  in the above calibrations are:
 0.029, 0.014, and 0.002 respectively.\\
 
For the RRab stars:
\begin{equation}
log~T_{\rm eff} = + ( 0.032 \pm 0.006 )~\rm{[Fe/H]}  + ( 3.852 \pm 0.008 )
\end{equation}
\begin{equation}
M_V~= + ( 0.191 \pm 0.037 )~\rm{[Fe/H]} + ( 1.032 \pm 0.054 )  
\end{equation}

The uncertainties in $log T_{\rm eff}$ and $M_V$ are 0.004 and 0.031 respectively.

 The relationship between $log L/L_{\odot}$ and [Fe/H] for the RRc stars (eq. 15)
is very tight, and can be transformed into a $M_V$ - [Fe/H] relationship with the
bolometric correction. Adopting the $BC - \rm{[Fe/H]}$ relation of Sandage \& Cacciari (1990) and
 $M_{bol,\odot}= 4.75$, we derive the relation:

\begin{equation}
 M_V= + ( 0.22 \pm 0.03 ) ~{\rm [Fe/H]} ~+~ ( 0.86 \pm 0.05 )
\end{equation}

This relation, shown as a dashed line in the lower panel of Fig. 10,
has a small zero point difference with
the average relationship obtained from several methods by Chaboyer (1999):
  ~$M_V = ( 0.23 \pm 0.04 )~\rm{[Fe/H]} ~+~ ( 0.93 \pm 0.12 )$.
Eq. 19 can be used to estimate an average $M_V= 0.53 \pm 0.08$ at
 [Fe/H] =$-$1.50. This result is in good agreement with the  
weighted average of $M_V = 0.58 \pm 0.04$ obtained by Cacciari (2003) from various
 methods, or with the weighted average of seven clusters obtained by Chaboyer (1999)
  $M_V = 0.61 \pm 0.11$, both for $\rm{[Fe/H]} = -1.50$.
Thus, the zero point predicted by the luminosities of eq. 3 (Simon \& Clement 1993)
 is in agreement with the above independent estimates, but it is
 about 0.20 mag. too bright relative to the zero point predicted by the calibrations
  of eqs. 7 and 9 for the RRc and RRab stars (Kov\'acs 1998;
   Kov\'acs \& Jurcsic 1996).

We do not find the discontinuity in the  $M_V$ $-$ [Fe/H] relation claimed by
 Lee \& Carney (1999b) but find rather a smooth transition between OoI and OoII clusters.\\

The above equations represent general laws for the horizontal branch position and structure
 as a function of metallicity and seem to hold for globular clusters at  a wide range
  of galactocentric distances and metallicities. 

The trends given by eqs. (15), (18), and (19) agree with previous studies in that
 the RR Lyrae variables are more luminous (hence more evolved) in lower metalicity (hence older) clusters. 

The relationship between $M_V$ and [Fe/H] has been widely discussed in the literature.
 Some evidence of the non linearity of the $M_V$ $-$ [Fe/H] relationship has been offered
 from empirical (Caputo et al. 2000; Demarque et al. 2000) and theoretical arguments
(Cassisi et al. 1999; Ferraro et al. 1999; VandenBerg et al. 2000).

Of particular interest is the $log M/M_{\odot}$ vs. [Fe/H] relation for the RRc variables.
 It is evident that RRc
 stars in OoII clusters have systematically larger masses than in OoI clusters. 
However, Clement \& Rowe (2000), in a study of RRc variables in the clusters $\omega$~Cen, M55 and M3
 reached the opposite conclusion. They found an increase of the mass with luminosity
 within each of the Oosterhoff groups, but a discontinuity at the transition
 between the OoI and OoII clusters. They derived lower masses and higher luminosities
  for the OoII RRc variables in $\omega$~Cen and M55 than the OoI RRc variables in $\omega$~Cen and M3,
 and a similar behaviour for the RRab variables. It seems that this conclusion could have been
 strongly influenced by the low value of $log M/M_{\odot}$ for the cluster M55. However,
  our Figs. 9 and 10 (see also Tables 7 and 8), based on a larger number of clusters, suggest
 a continuous trend with metallicity, and do not support any significant discontinuity between
 OoI and OoII variables.  
  
The trends in Figs. 9 and 10 and the related equations support the general scheme that
 OoII clusters are older and less metallic than OoI clusters, with their RR Lyrae
  being more massive, more luminous and having longer periods. This would be
expected if the mass loss at the tip of the Red Giant Branch is diminished for smaller
metallicities; thus in older clusters, at the time of reaching the HB stage, the stars turn into
 slightly more massive RR Lyrae
variables with lower $T_{\rm eff}$ but larger luminosity,
as expected from models of HB stars.
 This is consistent with the hypothesis that the origin of the
Oosterhoff dichotomy is age; in OoII the stars are more massive, older and more evolved off the HB.
The conclusion that on average metal poor (OoII group) are older than the richer clusters has been made
 in different studies (e.g.: Lee \& Carney 1999b).

\section{Conclusions}

Physical parameters of astrophysical relevance have been derived for the RR Lyrae stars in M2
using the Fourier decomposition of their light curves. 
The estimates of the stellar radii for RRab stars
from the above approach are about 5\% smaller than radii derived from Period-Radius-Metallicity 
relations obtained from nonlinear convective models, i.e.  
a completely different set of pulsational models and approach. The agreement for the RRc is 
generally good but the dispersion is markedly larger.
   
Previously recognized trends between the cluster mean values of $log~(M/M_{\odot})$, $log~(L/L_{\odot})$, 
 $log T_{\rm eff}$ and $M_V$ against [Fe/H], obtained from RRc stars, have been established
for a sample of 7 OoII and 5 OoI clusters. We extended the study of the
 trends for RRab stars in 4 OoII and 5 OoI clusters.

Our analysis, based on a larger data set than previous studies, allows one
to quantify the relations (see eqs. 14-19 and Figs.
 10 and 11). In particular, much has been discussed about the relation between
 luminosity (or  $M_V$) and metallicity for the RR Lyrae variables. Our eqs. 15, 17, and 19 are in very good
  agreement with the slopes of $M_V - [Fe/H]$ relations solidly determined 
and available in the literature. The zero points indicate however 
that the luminosities for the RRc and RRab stars, as calculated from eq. 3 (Simon \& Clement 1993) and
 those derived from eqs. 7 and 9 (Kov\'acs 1998; Kov\'acs \& Jurcsic 1996) respectively,
 differ by the equivalent to 0.2 mag. This suggests these calibrations need to be revised.
 
RR Lyrae Fourier light curve decomposition produces a zero point of the 
RR Lyrae distance scale consistent with those from other methods. When 
the light curve decomposition technique is used on families of OoI and OoII globular 
clusters, it provides physical parameters, as a function of the metal content of the cluster, that 
are consistent with general notions of stellar evolution in the HB and of early stages of galactic formation.

\section*{Acknowledgments}

The authors wish to thank S. Morgan for a careful reading of a first draft of the manuscript.
We thanks the referee to have pointed Bono et al. (1995) paper to our attention.
 AAF acknowledges support from DGAPA-UNAM grant through project IN110102 and is
  thankful to the Universidad de La Laguna and the Instituto de Astrof\1sica de
  Canarias (Spain) for hospitality. EP acknowledges support from Progetto INAF 39/2005.
   We thank Juan Carlos Yustis for assistance in manouvreing Figs. 5 and 6. This
    work has made a large use of the SIMBAD and ADS services, for which we are thankful.

\end{document}